\documentclass[twocolumn,dvipsnames]{aastex62}

\usepackage{framed}

\def\dim#1{{\rm\,#1}}

\submitjournal{ApJ}

\shorttitle{}
\shortauthors{}

\begin{document}

\title{Do Mini-halos Affect Cosmic Reionization?}

\correspondingauthor{Nickolay Y.\ Gnedin}
\email{gnedin@fnal.gov}

\author{Nickolay Y.\ Gnedin}
\affiliation{Fermi National Accelerator Laboratory;
Batavia, IL 60510, USA}
\affiliation{Kavli Institute for Cosmological Physics;
The University of Chicago;
Chicago, IL 60637 USA}
\affiliation{Department of Astronomy \& Astrophysics; 
The University of Chicago; 
Chicago, IL 60637 USA}

\begin{abstract}
The role of unresolved structures ("mini-halos") in determining the consumption of ionizing photons during cosmic reionization remains an unsolved problem in modeling cosmic reionization, despite recent extensive studies with small-box high-resolution simulations by Park et al.\ and Chan et al., because the small-box studies are not able to fully sample all environments. In this paper these simulations are combined with large-box simulations from the "Cosmic Reionization On Computers" (CROC) project, allowing one to account for the full range of environments and to produce an estimate for the number of recombinations per hydrogen atom that are missed in large-scale simulations like CROC or Thesan. I find that recombinations in unresolved mini-halos are completely negligible compared to recombinations produced in large-scale cosmic structures and inside more massive, fully resolved halos. Since both Park et al.\ and Chan et al.\ studies have severe limitations, the conclusions of this paper may need to be verified with more representative sets of small-box high-resolution simulations.
\end{abstract}


\section{Introduction}

The consumption of ionizing photons in unresolved structures (commonly dubbed "mini-halos") during cosmic reionization has been a small but active area of research since the first attempts to model the reionization process numerically. In the pioneering study, \citet{Haiman2001} estimated that numerical simulations that do not resolve mini-halos may underestimate the required number of ionizing photos by factors as large as 2 to 10. Later studies produced wildly varied estimates for this number, with some confirming it \citep{Shapiro2004,Iliev2005a,Iliev2005b,Emberson2013} and others finding significantly lower values \citep{Park2016,DAloisio2020,Chan2023}. This apparent inconsistency was, at least in part, due to variations in what counted as "unresolved". 

The underlying physics of how mini-halos affect reionization is well understood: (a) ionizing photons are consumed to ionize the dense gas inside them; (b) mini-halos are evaporated after being ionized since their virial temperatures are below the typical temperatures of photo-ionized gas (this is the definition of a "mini-halo"); (c) their dense gas recombines during evaporation, requiring additional ionizing photons to ionize it again. If ionizing radiation is intense, a significant number of ionizing photons can be absorbed during the evaporation stage, while the gas density is still high enough. If, however, ionization proceeds slowly, just one ionizing photon per baryon may be sufficient to raise the gas temperature above the virial temperature of a mini-halo, and hydrodynamics does the rest (M.\ Haehnelt, private communication).

The challenge in evaluating the effect of mini-halos on cosmic reionization accurately is the very large mass scales required. While simulations that model mini-halo evaporation typically have box sizes of hundreds of kpc, modeling the overall process of cosmic reionization requires simulation volumes hundreds of times larger \citep{Iliev2014}. Hence, no single simulation can resolve this question yet, and combining large-scale and small-scale models is needed for accounting for mini-halos. To the best of my knowledge, the first time this approach was implemented by \citet{Ciardi2006}, who used an analytical small-scale model and found a moderate, but not negligible effect. Similar studies were undertaken later only a handful of times \citep{Raicevic2011,Sobacchi2014,Mao2020,Park2023}.

In the last half a decade two major advances took place that warrant a revision to the question of photon consumption in mini-halos. First, large-scale, fully coupled simulation of reionization finally reached super-100-Mpc scales and $2000^3$ mass dynamic range - the flagship examples of such simulations are CROC\citep{Gnedin2014}, Coda \citep{coda1,coda2}, and Thesan \citep{thesan}. All these simulation projects reach similar resolutions, so the concept of "unresolved" is well-defined for them. Second, two larger parameter studies for photo-evaporation of mini-halos became available \citep{Park2016,Chan2023}. The goal of this paper is to combine large box simulations from CROC with \citet{Chan2023} and \citet{Park2016} models to evaluate the effect of unresolved small-scale structure on CROC reionization histories.

\section{Methodology}

Largest CROC simulations modeled reionization in boxes of $80h^{-1} \approx 117\dim{cMpc}$ on a side, with adaptive spatial resolution reaching 100 physical parsecs inside modeled galaxies but never exceeding $117/2048\dim{cMpc} \approx 57\dim{ckpc}$. One of the saved data products of CROC simulations is $1024^3$ uniform grids covering the whole computational domain. These grids are frequently sampled in time, allowing accurate time integration, and each cell in these grids has a size of at least two resolution elements, hence offering an independently resolved piece of spatial information with a size of $114\dim{ckpc}$.

In comparison, \citet{Chan2023} simulations were performed in $800\dim{ckpc}$ boxes, hence covering almost exactly a block of $7^3$ cells from CROC uniform grids. Those simulations modeled the effects of instantaneous reionization down to, at least, $1.6\dim{ckpc}$ (their mean inter-particle separation) with the maximum resolution set by the Plummer softening length of $0.1\dim{ckpc}$. The two main parameters that specify instantaneous reionization in \citet{Chan2023} simulations are the redshift of reionization $z_i$ and the global photoionization rate $\Gamma$. For a set of these two parameters, \citet{Chan2023} computed the number of recombinations per hydrogen atom $N_{\rm REC}/N_{\rm H}$ in the simulation volume as a function of time, and the ratio of that number to the number of recombinations in a uniform medium with the same median temperature, $N_{\rm REC}/N_{\rm UNI}$.

Their results for the latter can be accurately fitted with the following simple relation,
\begin{equation}
    \frac{N_{\rm REC}}{N_{\rm UNI}} = 1 + 5\psi(\Gamma)\left(\frac{1+z_i}{9}\right)^{-3.8}t_{100}^{-0.55},
    \label{eq:fit}
\end{equation}
where $t_{100}$ is the time since the instant of reionization in units of $100\dim{Myr}$. This fit becomes inaccurate for $t_{100}<1$.

The dependence on $\Gamma$ in \citet{Chan2023} results is somewhat non-trivial. One can fit them with a power-law,
\begin{equation}
    \psi_{\rm PL}(\Gamma) = \left(\frac{\Gamma_{-12}}{0.3}\right)^{0.3},
    \label{eq:psip}
\end{equation}
where $\Gamma_{-12} \equiv \Gamma/(10^{-12}\dim{s^{-1}})$, but the fit is not good enough. A more complex, double-power-law fit is significantly better, and I adopt it as a fiducial form,
\begin{equation}
    \psi_{\rm FID}(\Gamma) = 0.87055\times\left(\frac{\Gamma_{-12}}{0.15}\right)^\alpha,
    \label{eq:psif}
\end{equation}
where $\alpha = 0.2$ for $\Gamma_{-12}>0.15$ and $\alpha=0.32$ for smaller photoionization rates. Finally, one would expect the dependence on $\Gamma$ eventually saturate - when most of the gas is ionized, decreasing the neutral fraction from, say, $10^{-2}$ to $10^{-3}$ does not change the recombination rate, while the photoionization rate increases by a factor of 10. The following form includes saturation and provides an extremely accurate fit to the \citet{Chan2023} results:
\begin{equation}
    \psi_{\rm SAT}(\Gamma) = \frac{2.25x}{1+1.25x}, ~~~  x = \left(\frac{\Gamma_{-12}}{0.3}\right)^{0.5}.
    \label{eq:psis}
\end{equation}

\begin{figure}[t]
\centering
\includegraphics[width=\columnwidth]{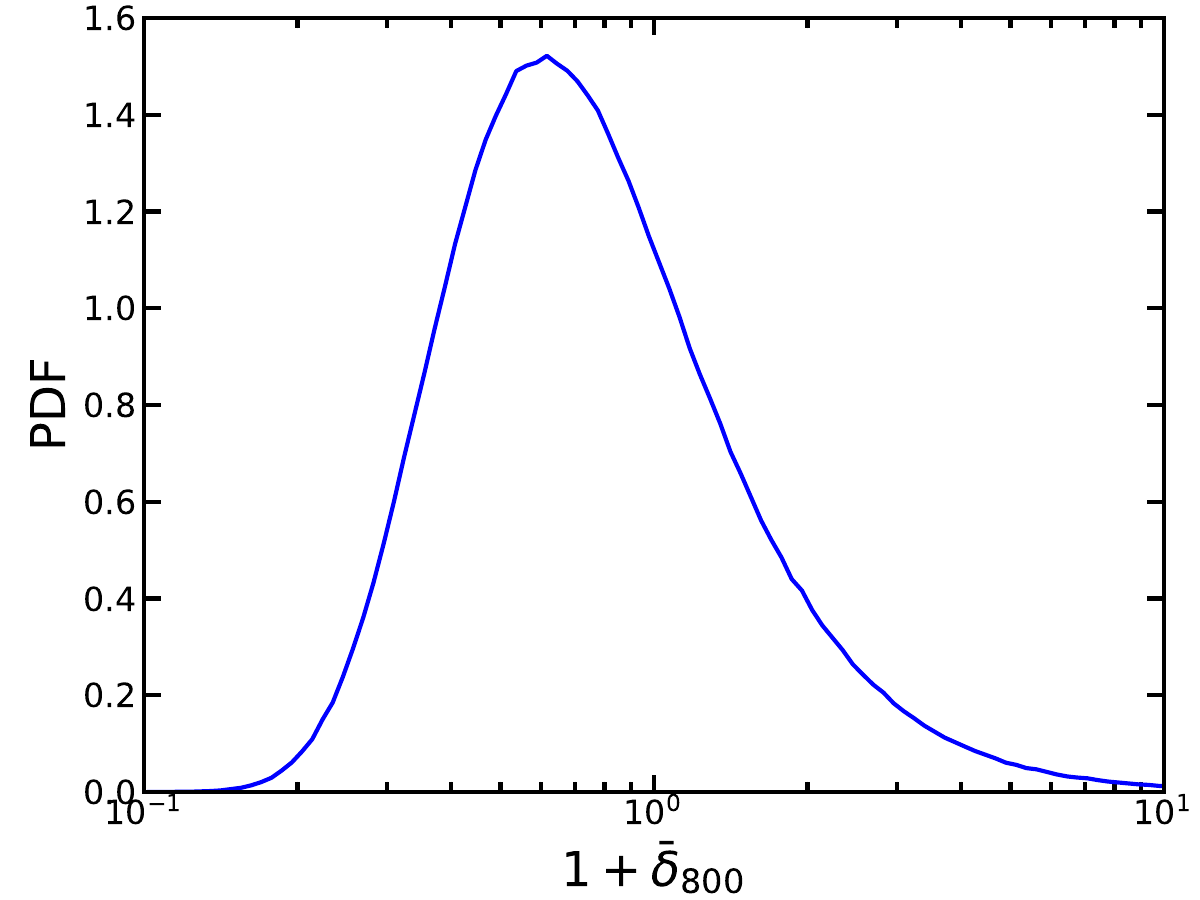}
\caption{The PDF of mean densities in $800\dim{ckpc}$ boxes at $z=5$. The non-negligible spread in densities is not accounted for in \citet{Chan2023} simulations but can be included using \citet{Park2016} simulations.\label{fig:pdf}}
\end{figure}

A major limitation of \citet{Chan2023} simulations is that they all adopt the mean cosmic density as the mean density in the simulation box, and hence do not account for the density variation in $800\dim{ckpc}$ boxes. That variation is shown in Figure \ref{fig:pdf} at $z=5$. If not accounted for, it will result in a significantly lower value for the number of recombinations. Fortunately, \citet{Park2016} performed an analogous simulation set in smaller, $300\dim{ckpc}$ boxes. \citet{Park2016} simulations used unrealistically large values for $\Gamma$ and continue for only $150\sim{Myr}$, which makes them less suitable for my purpose (and, as I show below, they substantially under-count recombinations). Fortunately, \citet{Park2016} did explore several boxes with mean densities deviating from the cosmic mean. Fitting the dependence on the mean density in these boxes, I find a highly expected result,
\[
    \frac{N_{\rm REC}}{N_{\rm H}} \propto (1+\bar{\delta})^2
\]
and
\[
    \frac{N_{\rm UNI}}{N_{\rm H}} \propto (1+\bar{\delta}).
\]
Notice, that \citet{Park2016} fitted the difference between $N_{\rm REC}/N_{\rm H}$ and $N_{\rm UNI}/N_{\rm H}$ and found a slightly steeper dependence, $\propto (1+\bar{\delta})^{2.5}$. This fit is 3 times worse (in its $p-value$)  than the two dependencies above, and obviously cannot extrapolate beyond their highest value for $\bar{\delta}=0.6$. Nevertheless, I show their exact results later as well.

The final model for the number of recombinations unresolved in $800\dim{ckpc}$ boxes is thus obtained by multiplying equation (\ref{eq:fit}) by $(1+\bar{\delta}_{800})$, where $\bar{\delta}_{800}$ is the average overdensity in $800\dim{ckpc}$ volume. I call this model "Chan-Park", as it combines both small-scale simulation sets.

In CROC simulations not every $800\dim{ckpc}$ block is reionized instantly. For example, at $z=7$ such a block has a length of $100\dim{pkpc}$. A typical cosmological ionization front, moving at $3000\dim{km/s}$, crosses this region in some $30\dim{Myr}$. Hence, one needs to come up with a working definition of reionization for such a region. For the sake of simplicity, I call such a block reionized when its mass-weighted neutral hydrogen fraction falls below a given threshold, $x_i$. 

To use the fitting formula (\ref{eq:fit}), one also needs to define the instantaneous photo-ionization rate $\Gamma$. Unfortunately, the uniform grid data from CROC simulations does not include that quantity, and saved full simulation snapshots are not sampled frequently enough for accurate time integration. I, thus, estimate the photo-ionization rate from the instantaneous reionization model, $x_{\rm HI}(t) = \exp(-\Gamma(t-t_i))$, where $t_i$ is the moment of reionization. Using two distinct moments $t_a$ and $t_b$, one can obtain an estimate of $\Gamma$ as
\[
    \Gamma \approx -\frac{\log(x_a)-\log(x_b)}{t_a-t_b}.
\]
Such an estimate includes a numerical error, so to account for it I use 3 different moments $t_1$, $t_2$, $t_3$ that corresponds to $\langle x_{\rm HI}\rangle_V=0.1, 0.03, 0.01$, and compute 2 estimates for $\Gamma$ with $(a,b)=(1,2)$ and $(a,b)=(2,3)$. Using the two estimates, I can compute the mean estimate and its error.

Hence, the whole computational procedure is as follows: for each $7^3$ block in a $1024^3$ uniform grid (the last 2 cells along each axis are ignored, so only $7\times146=1022$ cells used) I compute the estimate for the total number of recombinations $N_{\rm REC}$ as the product of the fitting formula (\ref{eq:fit}) and the number of recombinations in an equivalent uniform medium with the density, neutral fraction, and temperature averaged over the $7^3$ from a CROC simulation. In order to check the dependence on the numerical parameter $x_i$, I use $x_i=0.3, 0.1, 0.03$.

\begin{figure*}[t]
\centering
\includegraphics[width=\columnwidth]{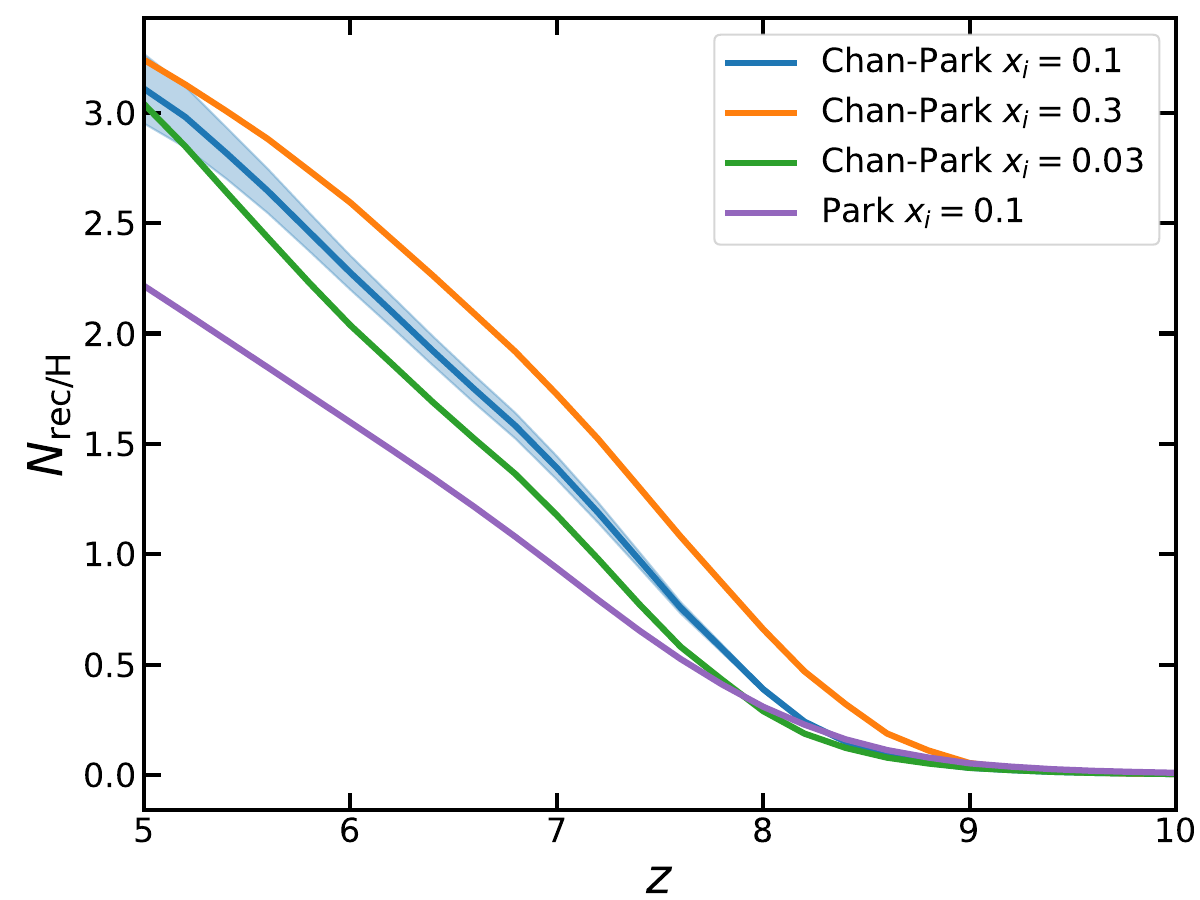}
\includegraphics[width=\columnwidth]{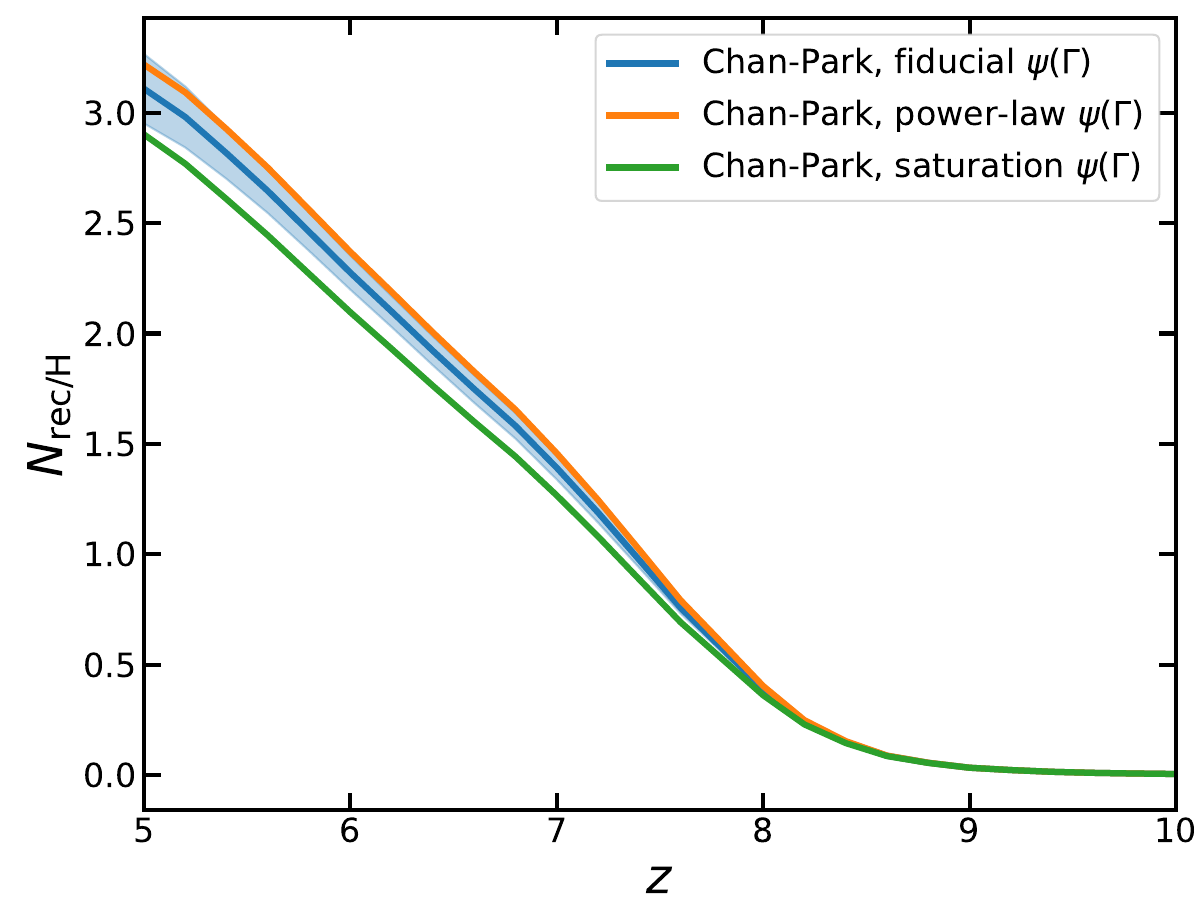}
\caption{Left: the total number of recombinations per hydrogen atom for the three choices of the instant reionization threshold in the full $117\dim{Mpc}$ CROC box. The translucent band shows the error in the mean as measured from 2 different numerical estimates of the photo-ionization rate $\Gamma$. The purple line shows the $x_i=0.1$ case for \citet{Park2016} (their Equation 22) applied to CROC blocks of $3^3$ cells (the best match to their $300\dim{ckpc}$ boxes). Right: the same quantity for 3 different fits to the $\Gamma$ dependence, Equations (\ref{eq:psip}-\ref{eq:psis}). \label{fig:met}}
\end{figure*}

\begin{figure*}[ht]
\centering
\includegraphics[width=\columnwidth]{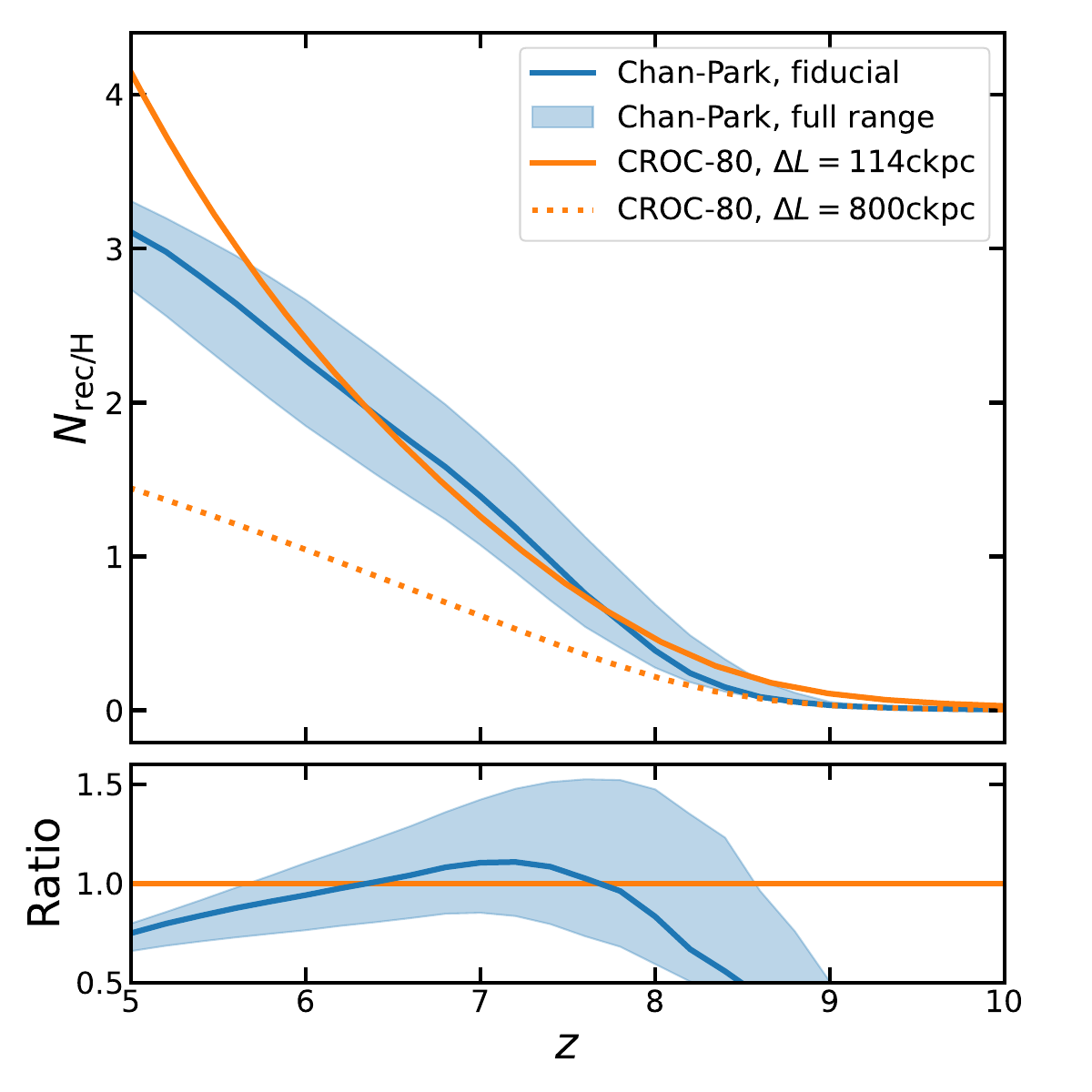}
\includegraphics[width=\columnwidth]{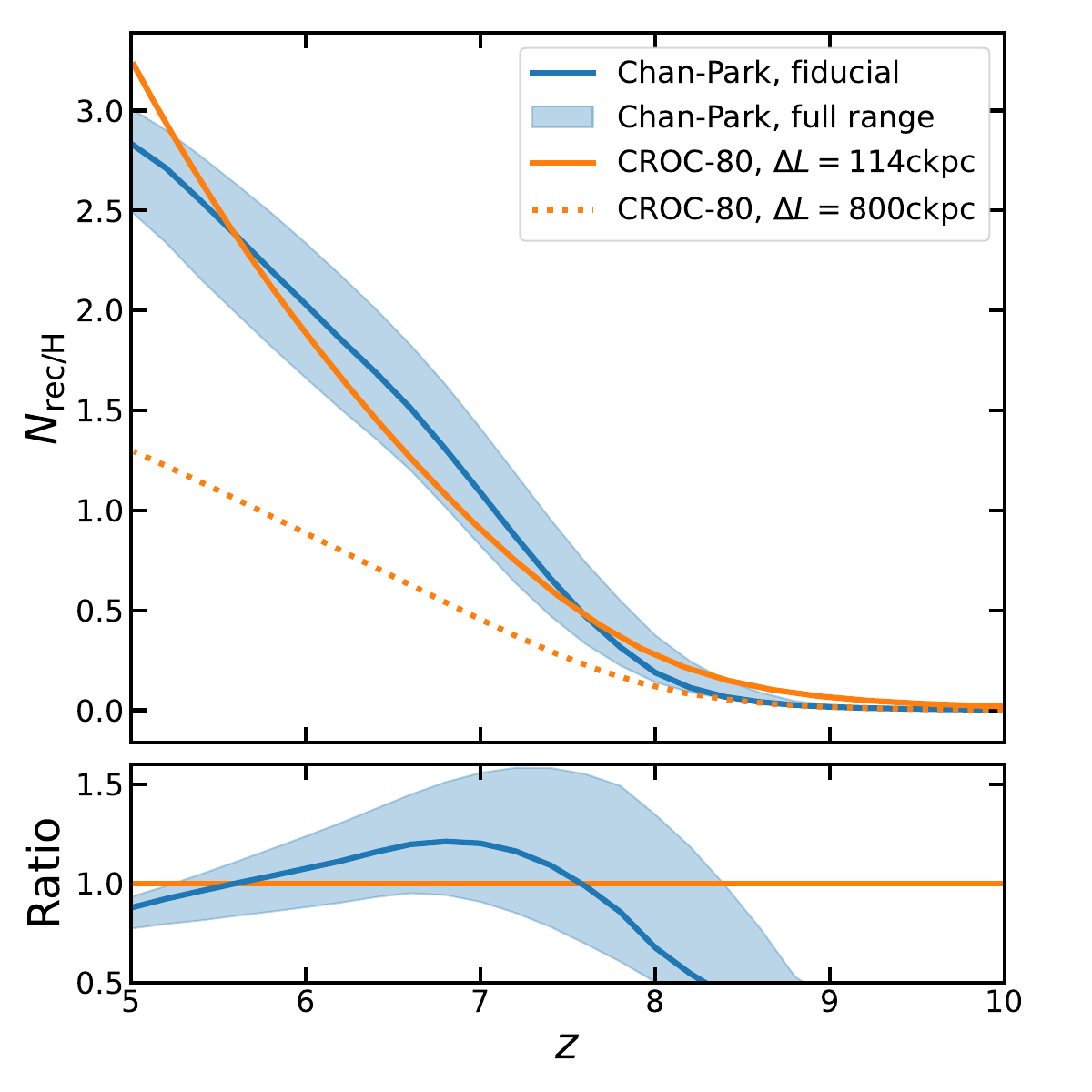}
\caption{The total number of recombinations per hydrogen atom for two CROC simulations: fiducial "earlier reionization" (left) and "later reionization" (right). Blue lines with bands show the fiducial Chan-Park model with the full range (from minimum to maximum) of uncertainties due to methodological choices. Orange solid and dotted lines show $N_{\rm REC}/N_{\rm H}$ and $N_{\rm UNI}/N_{\rm H}$ from CROC simulations, which can be interpreted as the total number of recombination per hydrogen atom from uniform blocs of $\Delta L=114\dim{ckpc}$ and $\Delta L=800\dim{ckpc}$ respectively.
\label{fig:res}}
\end{figure*}

Figure \ref{fig:met} shows the total (i.e.\ the cosmic mean) number of recombinations per hydrogen atom in the fiducial CROC simulation (described in more detail below). Two panels show the effects of various methodological choices. I elect the $x_i=0.1$ with the double-power-law fit for $\psi(\Gamma)$ (eq.\ \ref{eq:psif}) as "fiducial", and I also record the maximum and the minimum models to demonstrate the limitations of the chosen approach. The left panel of the figure also shows the calculation using Equation (22) from \citet{Park2016}, with their steeper density dependence, as applied to $3^3$ blocks from the CROC simulation. \citet{Park2016} fit under-counts the number of recombinations in comparison to the full Chan-Park model, primarily because their simulations do not extend beyond $150\dim{Myr}$. The slope of $-0.55$ in the time dependence of \citet{Chan2023} simulation demonstrates that there is still a significant number of recombinations taking place well after $150\dim{Myr}$.

\section{Results and Conclusions}

Fig.\ \ref{fig:met} would pretty much represent the results of this short paper if the CROC simulations had a resolution of $800\dim{ckpc}$. However, CROC simulations cover the whole computational domain with cells of at most $57\dim{ckpc}$ on a side, and much smaller in highly resolved regions. With the easily available uniform grid data I can compute both the number of recombinations per hydrogen atom ion the full $1024^3$ CROC uniform grid, and for the same grid with all physical quantities smoothed in $800\dim{ckpc}$ blocks ($7^3$ cells). 

I do this for two CROC simulations from \citet{Gnedin2022}. Both of these simulations have sizes of $80h^{-1}\approx 117\dim{cMpc}$. The first of the two, which I call "fiducial" or "earlier reionization", is referred to as "DC=0" in \citet{Gnedin2022}. This simulation provides a reasonable match to the observed distribution of mean opacities in $50h^{-1}\dim{Mpc}$ skewers from \citet{Becker2015} and \citet{Bosman2018}, but fails to match the observed distribution of "dark gaps" (regions in quasar absorption spectra without significant transmission) from \citet{Zhu2021}. The second simulation, which I call "later reionization", is referred to as "DC=-1" in \citet{Gnedin2022}. That simulation matches the distribution of dark gaps very well but fails to match the distribution of mean opacities. Thus, the two simulations each match one of the two key observables and fail to match the other. Unfortunately, this is currently the state-of-the-art in reionization modeling \citep{Gnedin2022b}.

Figure \ref{fig:res} now shows the main result of this paper - the total number of recombinations per hydrogen atom estimated with the Chan-Park model (both the fiducial approach and the full range due to methodological choices described above) as compared with the actual number of recombinations computed in two CROC simulations. In both cases, the fiducial approach finds barely any recombinations from the structures unresolved in CROC. In the most extreme scenario, there may be up to 50\% more recombinations produced that are captured in CROC $1024^3$ uni-grids.

However, the actual coarsest resolution of $80h^{-1}$ Mpc CROC boxes is twice higher, the equivalent of a $2048^3$ uni-grid (with the adaptive resolution even higher in adaptively refined regions). Because of the limitations of the available storage, only $1024^3$ uni-grids were produced for $80h^{-1}$ Mpc CROC boxes, and the same size grids were also produced for 6 smaller, $40h^{-1}$ Mpc CROC boxes. Hence, these smaller boxes have twice higher resolution of their $1024^3$ uni-grids. They can be used to check the effect of the uni-grid resolution on the results of this paper. 

\begin{figure}[t]
\centering
\includegraphics[width=\columnwidth]{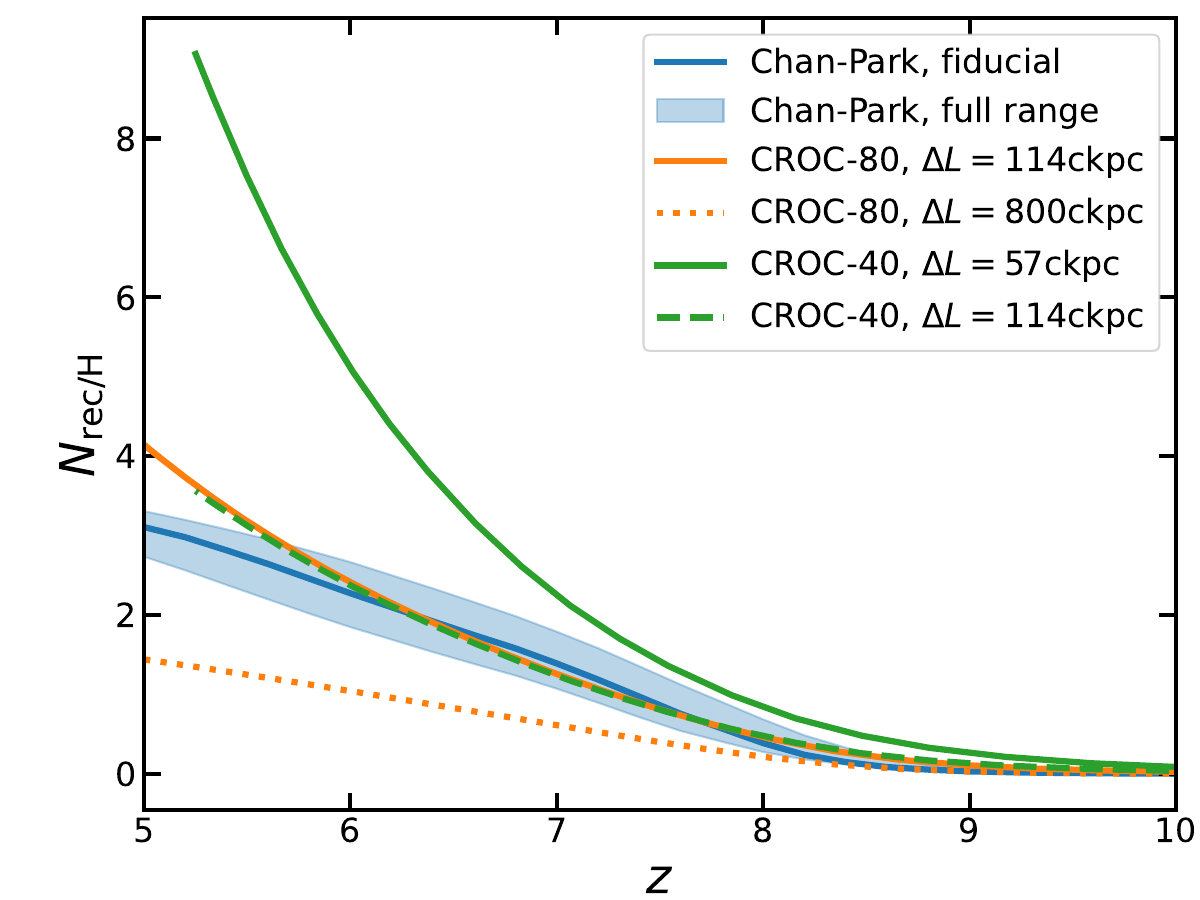}
\caption{The same as the left panel of Fig.\ \ref{fig:res}, and now also showing recombinations from 2 smaller, $40h^{-1}$ Mpc boxes with the same size, $1024^3$ uni-grids, that offer twice higher uniform spatial resolution. With additional resolution, the contribution of unresolved strictures becomes negligibly small.
\label{fig:res2}}
\end{figure}

Two $40h^{-1}$ Mpc CROC boxes happen to have together almost the same number of recombinations per hydrogen atom as the fiducial $80h^{-1}$ Mpc box when averaged with the same uni-grid resolution ($512^3$, $\Delta L=114\dim{ckpc}$). The total number of recombinations per hydrogen atom from these two $40h^{-1}$ Mpc boxes without averaging (i.e.\ in original $1024^3$, $\Delta L=57\dim{ckpc}$ blocks) is shown in Fig.\ \ref{fig:res2}, together with all other lines from the left panel of Fig.\ \ref{fig:res}. The additional factor of 2 increase in resolution significantly increases the total number of recombinations per hydrogen atom, totally drowning any possible contribution from mini-halos. This increase should not be over-interpreted - as the resolution increases, the contributions from recombinations from CGM and ISM of galaxies increase, but with tens of ckpc resolution, these contributions are not captured accurately. In addition, at small enough scales, some of the recombinations are balanced by collisional ionizations, which are not accounted for in this work. The increase in the number of recombinations between 114 and 57 ckpc resolution simply implies that at such scales the counting of recombinations becomes complex and ambiguous, with recombinations from CGM, ISM, and those balanced by collisional ionizations requiring special treatment.

In conclusion, it appears that the mass and spatial resolution of the current generation of large reionization simulations, such as CROC or Thesan, is sufficient to account for all recombinations, and that unresolved recombinations in mini-halos are negligible for all practical purposes. This conclusion is based on \citet{Park2016} and \citep{Chan2023} simulations. These simulation sets can be improved further by, for example, adding the density dependence to \citep{Chan2023} simulations and improving the sampling of the photo-ionization rate dependence. If such extensions of the existing small box simulations become available, the conclusions of this paper will need to be revised.

\acknowledgments

I am grateful to Tsang Keung Chan and Hyunbae Park for constructive comments that significantly improved the original manuscript. This work was supported in part by the NASA Theoretical and Computational Astrophysics Network (TCAN) grant 80NSSC21K0271. This manuscript has also been co-authored by Fermi Research Alliance, LLC under Contract No. DE-AC02-07CH11359 with the U.S. Department of Energy, Office of Science, Office of High Energy Physics. This work used resources from the Argonne Leadership Computing Facility, which is a DOE Office of Science User Facility supported under Contract DE-AC02-06CH11357. An award for computer time was provided by the Innovative and Novel Computational Impact on Theory and Experiment (INCITE) program. This research is also part of the Blue Waters sustained-petascale computing project, which is supported by the National Science Foundation (awards OCI-0725070 and ACI-1238993) and the state of Illinois. Blue Waters is a joint effort of the University of Illinois at Urbana-Champaign and its National Center for Supercomputing Applications. This work was completed in part with resources provided by the University of Chicago Research Computing Center.

\bibliographystyle{aasjournal}
\bibliography{main}

\end{document}